\documentclass[aps,prb,reprint,nofootinbib,superscriptaddress,twocolumn,showpacs,showkeys,longbibliography,amsmath,amssymb]{revtex4-1}
\usepackage{graphicx}% Include figure files
\usepackage{dcolumn}% Align table columns on decimal point
\usepackage{bm}% bold math
\usepackage{braket}
\usepackage{subfigure}
\usepackage[colorlinks,bookmarks=false,citecolor=blue,linkcolor=red,urlcolor=blue]{hyperref}
\usepackage[english]{babel}
\usepackage{changes}

\begin{document}
%\begin{bibunit}

\title{Topological magnons in Kitaev magnets with finite Dzyaloshinskii-Moriya interaction at high field}
\author{Kangkang Li}
\email{physeeks@163.com}
\affiliation{Department of Physics, Zhejiang Normal University, Jinhua, 321004, China}

\begin{abstract}
There have been intensive studies on Kitaev materials for the sake of the realization of exotic states such as quantum spin liquid and topological orders. In realistic materials, the Kitaev interaction may coexist with the Dzyaloshinskii-Moriya interaction (DMI), and it is a challenge to distinguish their magnitudes separately. Here, we study the topological magnon excitations and related thermal Hall conductivity of kagome magnet exhibiting Heisenberg, Kitaev and DM interactions exposed to a magnetic field. In a strong magnetic field perpendicular to the plane of the lattice ([111] direction) that bring the system into the fully polarized paramagnetic phase, we find that the magnon bands carry nontrivial Chern numbers in the full region of the phase diagram. Furthermore, there are phase transitions relate two topological phases with opposite Chern numbers, which lead to the sign changes of the thermal Hall conductivity. In the phase with negative thermal conductivity, the Kitaev interaction is relatively large and the width of the phase increases with the strength of DMI. Hence the study here will contribute to the understanding of related compounds.
\end{abstract}
\maketitle

\section{Introduction}\label{sectoin1}
In recent years, quantum materials with bond-dependent anisotropic Kitaev spin interactions have been the subject of much experimental and theoretical studies, because there are frustrations from the competing exchange couplings coexisting with the geometric frustration of the underlying lattices, which may lead to unusual magnetic orders as well as gapped and gapless spin liquids with fractional excitations\cite{Rou,Lee,Nasu,Taka,Jah,Jah2,Kishi,Kim}. However, the Kitaev interaction is often accompanied by the Dzyaloshinskii-Moriya interaction (DMI)\cite{DM,DM2} in realistic materials, and it is a challenge to distinguish their magnitudes separately\cite{Gao,Zhangli}.

The Kitaev interaction also shows its exotic properties in magnonic side, for instance, it can realize topological magnon bands in various lattices such as honeycomb lattice\cite{Joshi,Chern,Mcc,Zhangli} and kagome lattice\cite{Kang}. However, the effect of the Kitaev interaction can be generally similar to that of the DMI, which can also induce topological magnons and has been already studied in lots of experimental and theoretical works\cite{Kat,Onose,Hir,Chi,Owerre,Zhang,Kang2,Kang3,Cao}. In some cases, the spin wave spectrum of the Heisenberg-Kitaev model can even reduce to that of the Heisenberg model with DMI, and then hosts the same topological property\cite{Mcc}. Therefore, the same magnon bands and topological property can be generated either by Kitaev interaction, DMI, or their combination. Fortunately, it was shown that one can distinguish whether a system is Kitaev interaction or DMI dominated by further investigating the magnonic transport properties, for example, in honeycomb Kitaev magnet\cite{Gao,Zhangli}.

\begin{figure}[bth]
      \centering
      \includegraphics[width=0.48\textwidth]{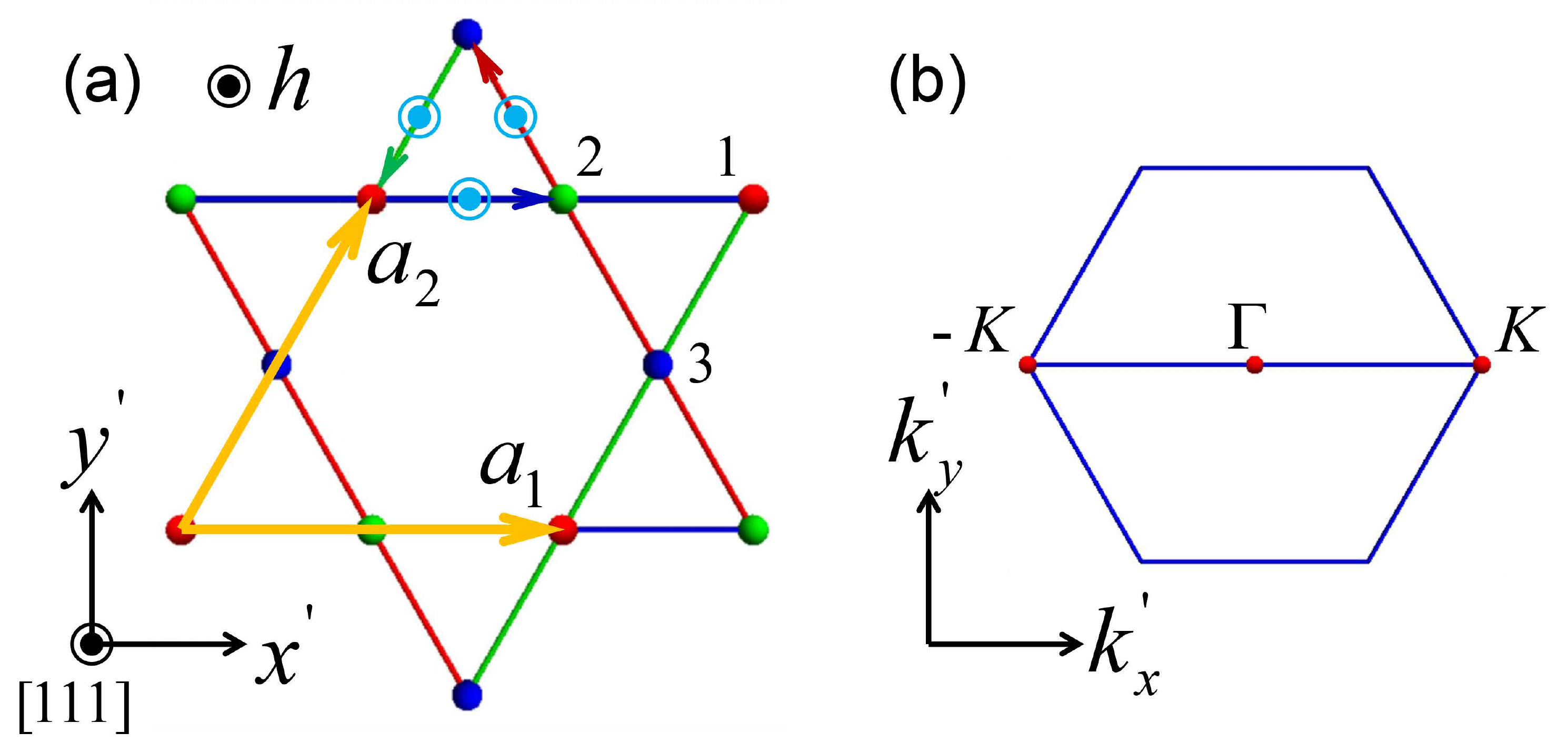}
      \caption{(a) The structure of the kagome lattice. There are three spins reside in a primitive cell, which are denoted by red, green and blue sites. Red, green and blue bonds between NN sites $(i,j)$ carry three distinct Kitaev couplings $S_i^xS_j^x$, $S_i^yS_j^y$ and $S_i^xS_j^x$, respectively. Meanwhile, there are isotropic Heisenberg couplings in all the NN bonds. The arrows in a triangle denote the coupling directions whose DM vectors are in the [111] direction. The field is also applied in the [111] direction. The kagome lattice sits on the (111) plane, and we define a new 2D frame $x^\prime y^\prime$ on the kagome lattice with basis vectors $\bm{a}_1=(1, 0)$ and $\bm{a}_2=(1,\sqrt{3})/2$. (b) The first Brillouin zone of the kagome lattice.}
      \label{Fig1}
\end{figure}

\begin{figure*}[tbh]
      \centering
      \includegraphics[width=0.9\textwidth]{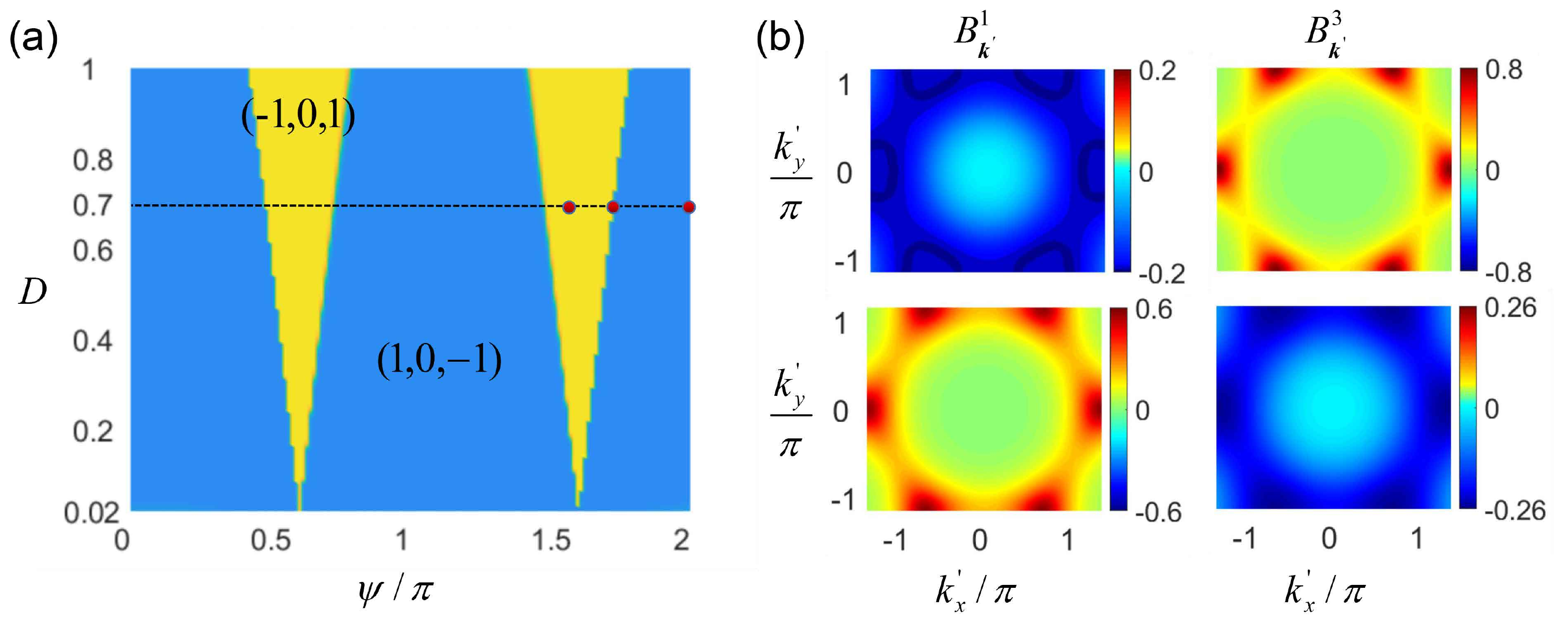}
      \caption{(a) The topological phase diagram of our model as a function of $\psi$ and $D$. The topological phases are characterized by the Chern numbers of magnon bands, and the phase diagram is filled by two phases $(1,0,-1)$ and $(-1,0,1)$. The three red points on the dashed line with $D=0.7$ reside in phase $(-1,0,1)$ with $\psi=1.57\pi$, at phase boundary with $\psi=1.726\pi$ and in phase $(1,0,-1)$ with $\psi=2\pi$, respectively. (b) The Berry curvatures of the first and third magnon bands for $\psi=1.57\pi$ (top panel) and $\psi=2\pi$ (bottom panel) with $D=0.7$.}
      \label{Fig2}
\end{figure*}

In this work, we study the topological magnon excitations and related thermal Hall conductivity of kagome magnet exhibiting Heisenberg, Kitaev and DM interactions exposed to a magnetic field. It is well known that the magnon bands of kagome ferromagnet with DMI carry nonzero Chern numbers\cite{Chi}. On the other hand, the kagome magnet with Kitaev interaction, whose ground state is canted ferromagnetic order\cite{HK1,HK2}, also hosts topological magnon excitations\cite{Kang}. Hence, one may wonder what would happen when there are DMI and Kitaev interaction in kagome magnet simultaneously. Since the Kitaev interaction does not support ferromagnetic order\cite{HK1,HK2}, we apply a magnetic field perpendicular to the plane of the lattice ([111] direction), which is strong enough to bring the system into fully polarized paramagnetic phase\cite{Mcc}. We find that the magnon bands carry nontrivial Chern numbers in the full region of the phase diagram. Furthermore, there are phase transitions relate two topological phases with opposite Chern numbers, which lead to the sign changes of the thermal Hall conductivity. In the phase with negative thermal conductivity, the Kitaev interaction is relatively large and the width of the phase increases with the strength of DMI. Thus the study here will contribute to the understanding of related compounds.

This paper is organized as follows. In section \ref{sectoin2}, we introduce the model and related methods. In section \ref{sectoin3}, we discuss the topological magnons. In section \ref{sectoin4}, we present the thermal Hall conductivity with sign change. Finally, a summary is given in section \ref{sectoin5}.

\section{Model and method}\label{sectoin2}
We consider interacting spins reside on the kagome lattice as shown in Fig. \ref{Fig1}(a). The model is described by the spin Hamiltonian
\begin{align}
H=&J\sum_{\langle ij\rangle}{\bf S}_i\cdot{\bf S}_j+K\sum_{\langle ij\rangle}S_i^{\gamma_{ij}}S_j^{\gamma_{ij}}\notag\\
&+D\sum_{\langle ij\rangle}\nu_{ij}\hat{{\bm z}^\prime}\cdot({\bf S}_i\times{\bf S}_j)-h\sum_i\hat{{\bm z}^\prime}\cdot{\bf S}_i,
\label{model}
\end{align}
where ${\bf S}_i$ and ${\bf S}_j$ are spin $S=1/2$ spins reside on the nearest-neighbor (NN) lattice sites, and $J$ and $K$ denote the Heisenberg and Kitaev exchange couplings, respectively. The Cartesian components $\gamma_{ij}$ equals $x$, $y$ or $z$, depending on the bond type as shown in Fig. \ref{Fig1}(a). We parameterize the Heisenberg and Kitaev terms by $J=\cos\psi,\quad K=\sin\psi,\quad \psi \in [0, 2\pi)$, with the energy unit $J^2+K^2=1$. The NN DMI is represented by the third term, where $\hat{{\bm z}^\prime}$ is the unit vector along [111] direction and $\nu_{ij}=\pm1$ correspond to the anticlockwise and clockwise directions of the NN couplings in a triangle, respectively. In kagome magnet the Kitaev interaction does not support ferromagnetic order\cite{HK1,HK2}, hence we add the final term of Zeeman coupling to a applied magnetic field along the [111] direction, which is strong enough to bring the system into fully polarized paramagnetic phase. We study a reasonable parameter range of $D\in[0, 1]$, and set $h=10$ (in unit of $S$) all through the work to ensure the stability of the system.

Further, the Holstein-Primakoff (HP) transformation\cite{HP} is employed to rewrite the Hamiltonian in terms of magnon creation and annihilation operators $a_i^\dagger$ and $a_i$. Under the linear spin wave approximation, we keep only the quadratic terms of the magnon operators. Then the Fourier transformation is performed to rewrite the Hamiltonian in momentum space with basis $\Psi_{\bm{k}^\prime}^\dagger=(a_{1\bm{k}^\prime}^\dagger, a_{2\bm{k}^\prime}^\dagger, a_{3\bm{k}^\prime}^\dagger, a_{1-\bm{k}^\prime}, a_{2-\bm{k}^\prime}, a_{3-\bm{k}^\prime})$, and we get the magnon Hamiltonian matrix $h(\bm{k}^{\prime})$ (see Appendix for details). Finally, the eigenvalues and eigenvectors are obtained by diagonalizing the dynamic matrix $I_-h(\bm{k}^{\prime})$ with $I_-=[(I,0),(0,-I)]$, $I$ as the $3\times3$ identity matrix.

\section{Topological magnons}\label{sectoin3}

\begin{figure*}[tbh]
      \centering
      \includegraphics[width=0.55\textwidth]{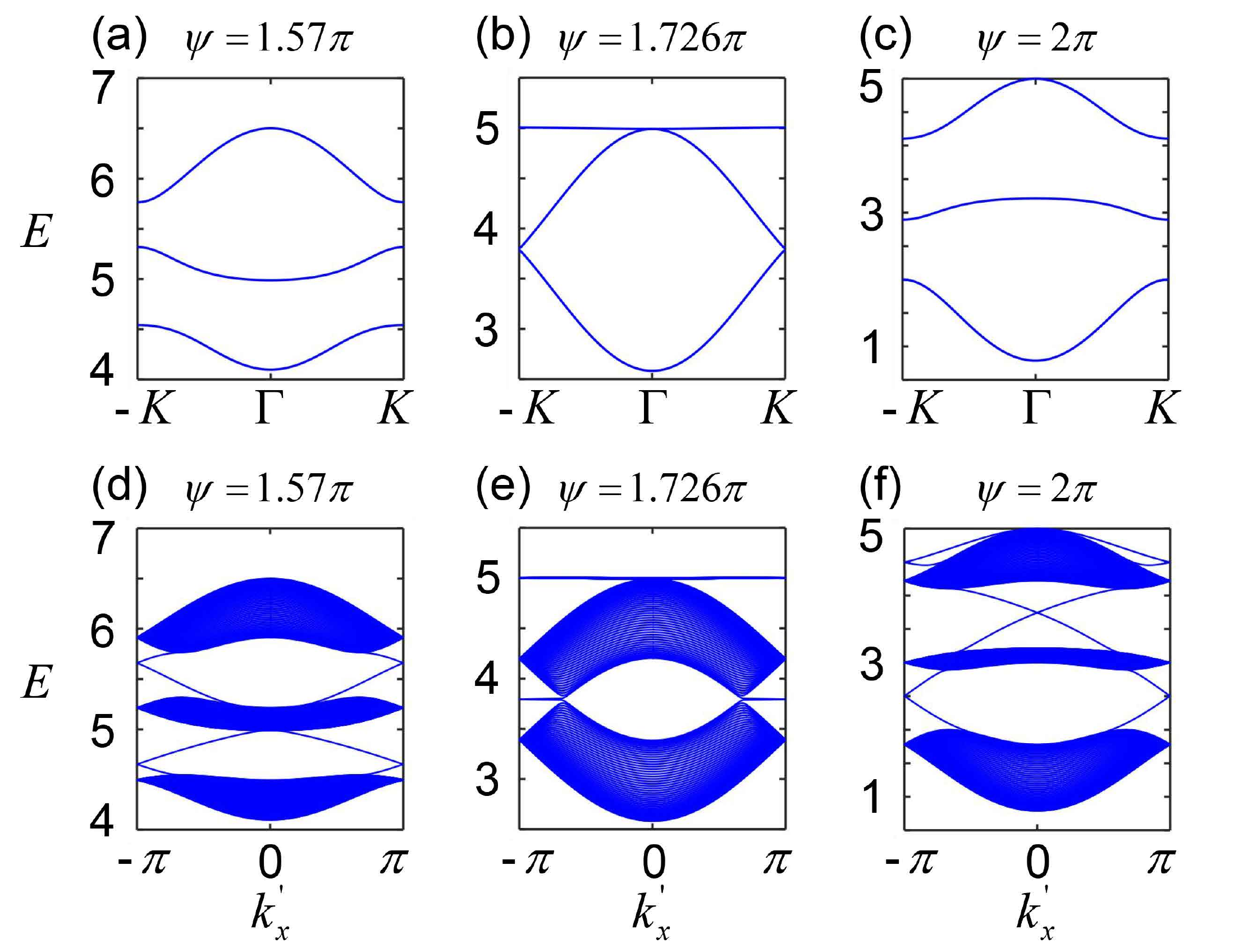}
      \caption{Top panel: magnon bands for (a) $\psi=1.57\pi$, (b) $\psi=1.726\pi$ and (c) $\psi=2\pi$ all with $D=0.7$, which are denoted in Fig. \ref{Fig2}(a) as red points. For the point (b) sits at the phase boundary, the bands are gapless and there are Dirac points at the $K$ points. Note that the top band in (b) has small dispersion. Bottom panel: the corresponding magnon bands in a strip geometry. There are in-gap edge modes in (d) and (f), and nondispersive edge modes connecting the Dirac points in (e).}
      \label{Fig3}
\end{figure*}

We characterize the topological property of the model by the Chern number\cite{Chern1,Chern2,Chern3} of the $n$th magnon band
\begin{align}
C_n=\frac{1}{2\pi}\int_{\rm BZ}{\rm d}k_x^{\prime}{\rm d}k_y^{\prime}B_{k_x^{\prime}k_y^{\prime}}^n,
\end{align}
with the Berry curvature of the $n$th band
\begin{align}
B_{k_x^{\prime}k_y^{\prime}}^n\!\!={\rm i}\!\!\sum_{n^\prime\ne n}\!\!\frac{\langle\phi_n|\frac{\partial h(\bm{k}^{\prime})}{\partial k_x^{\prime}}|\phi_{n^\prime}\rangle\langle\phi_{n^\prime}|\frac{\partial h(\bm{k}^{\prime})}{\partial k_y^{\prime}}|\phi_n\rangle\!\!-\!\!(k_x^{\prime}\leftrightarrow k_y^{\prime})}{(E_n-E_{n^\prime})^2},
\end{align}
where $E_n$ and $\phi_n$ are the eigenvalue and eigenvector of the $n$th band respectively.

The Heisenberg kagome magnet with only DMI or Kitaev interaction both host topological magnon excitations\cite{Chi,Kang}. The similar effects on magnon topology of the DMI and Kitaev interaction stem from their common origin, the spin-orbit coupling. However, due to their different formalisms of spin operators, there should also be competing effects. Here we find that the magnon bands of our model carry nontrivial Chern numbers in the full region of the phase diagram with the tuning of $D$ and $\psi$, as shown in Fig. \ref{Fig2}(a). We characterize the topological phase by $(C_1,C_2,C_3)$, and the phase diagram is filled by two phases $(1,0,-1)$ and $(-1,0,1)$. Note that we only show the parameter range $D\in[0.02,1]$ for the phase diagram, because there are complicated phases and phase transitions in the range $D\in[0,0.02]$ and the magnitude of the related thermal conductivity is nearly zero, and these results are deviated from the theme of this paper.

One may note that, the two separated regions of the phase $(-1,0,1)$ have nearly the same shape, and they may can be related to each other by $\psi+\pi$. However, it is not exactly the case. The mapping $\psi\rightarrow\psi+\pi$ will take $(J,K)$ to $(-J,-K)$ and the Zeeman coupling to the field will only give a energy shift of $h$ for the whole spectrum, then if we have the eigenvalue of the $n$th band $E_{n,\psi,D}(\bm{k}^\prime)=\widetilde{E}_{n}(\bm{k}^\prime)+h$, we must also have $E_{n,\psi+\pi,-D}(\bm{k}^\prime)=-\widetilde{E}_{n}(\bm{k}^\prime)+h$. Hence we have the relation $(E_{n,\psi,D}(\bm{k}^\prime)+E_{n,\psi+\pi,-D}(\bm{k}^\prime))/2=h$, which means $E_{n,\psi,D}(\bm{k}^\prime)$ and $E_{n,\psi+\pi,-D}(\bm{k}^\prime)$ are mirror reflections of each other about the energy axis $E=h$. Thus the closing and re-opening of the band gap, which denotes a topological phase transition, is simultaneous for the bands $E_{n,\psi,D}(\bm{k}^\prime)$ and $E_{n,\psi+\pi,-D}(\bm{k}^\prime)$. Therefore, if we have $E_{n,\psi+\pi,-D}(\bm{k}^\prime)=E_{n,\psi+\pi,D}(\bm{k}^\prime)$, the two separated regions of the phase $(-1,0,1)$ will be related to each other by $\psi+\pi$ and have exactly the same shape. However, we have checked that the sign change of $D$ will affect the band shape slightly and break the simultaneity of the related phase transitions. Consequently, we only have similar but not the same shape of the two regions of the phase $(-1,0,1)$. Now the same Chern numbers of the two regions can also be understood easily. The mirror reflection about a constant energy axis does not change the Chern numbers of the bands, however, it will change the order of the bands. If the bands $E_{n,\psi,D}(\bm{k}^\prime)$ carry Chern numbers $(-1,0,1)$, then the mirror-reflected bands $E_{n,\psi+\pi,-D}(\bm{k}^\prime)$ will carry Chern numbers $(1,0,-1)$. However, the sign change of $D$ will change the sign of the Chern number of every band. Thus the bands $E_{n,\psi+\pi,D}(\bm{k}^\prime)$ will carry Chern numbers $(-1,0,1)$, which are the same with that of the bands $E_{n,\psi,D}(\bm{k}^\prime)$.

To gain a deeper insight of the topological property, we calculate the Berry curvatures of the magnon bands for $\psi=1.57\pi$ and $\psi=2\pi$ both with $D=0.7$, which are in the phases $(-1,0,1)$ and $(1,0,-1)$ respectively, as shown in Fig. \ref{Fig2}(b). As for $\psi=1.57\pi$ with $D=0.7$, the Berry curvature of the first band is negative all over the Brillouin zone, and the integral of it gives Chern number $-1$. While for the third band, its Berry curvature is always positive and gives Chern number $1$. For $\psi=2\pi$ with $D=0.7$, the situation is reversed. The sign of the Berry curvature of the second band is momentum dependent for both cases, and they are not shown here.

\begin{figure*}[tbh]
      \centering
      \includegraphics[width=0.9\textwidth]{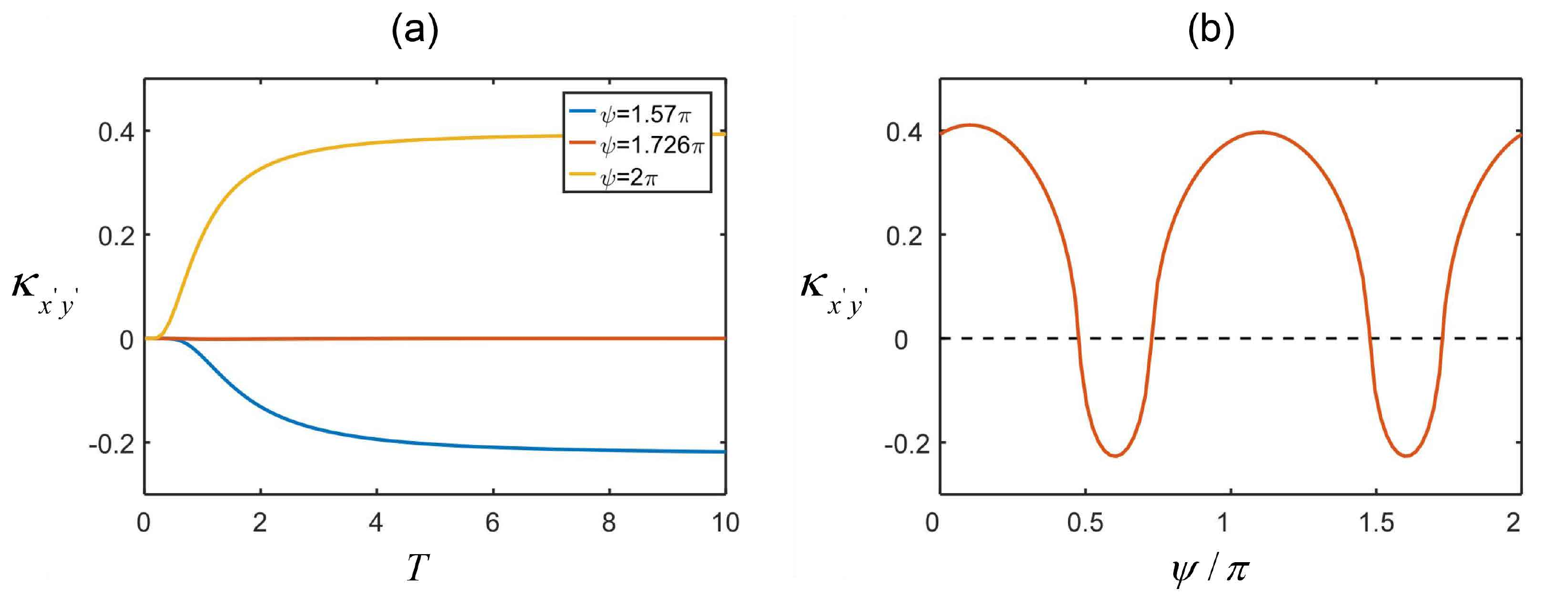}
      \caption{(a) The transverse thermal conductivity as function of temperature for $\psi=1.57\pi$, $\psi=1.726\pi$ and $\psi=2\pi$ all with $D=0.7$, which are denoted in Fig. \ref{Fig2}(a) as red points. The thermal conductivity are always negative and positive for the phases $(-1,0,1)$ and $(1,0,-1)$ respectively, and it is always zero for the phase boundary. (b) The transverse thermal conductivity as function of $\psi$ with $D=0.7$ and $T=10$. There are sign changes at the phase boundaries. Here we set $\hbar=k_B=1$.}
      \label{Fig4}
\end{figure*}

In Fig. \ref{Fig3} we show the bulk magnon bands and their corresponding band structures in a strip geometry for $\psi=1.57\pi$, $\psi=1.726\pi$ and $\psi=2\pi$ all with $D=0.7$, which are denoted in Fig. \ref{Fig2}(a) as red points. For the point sits at the phase boundary, the band structure is nearly the same with that of the Heisenberg kagome ferromagnet without DMI, the bands are gapless and there are Dirac points at the $K$ points\cite{Chi}, except that here the top band has small dispersion. Due to the bulk-edge correspondence\cite{Hat}, there are in-gap edge modes in the gapped band structures and nondispersive edge modes connecting the Dirac points.

\section{Transverse thermal Hall conductivity with sign change}\label{sectoin4}
The momentum independent sign structure of the Berry curvatures will lead to the sign change of the magnon thermal Hall conductivity, as we will show in this section. The transverse thermal Hall conductivity of magnon can be calculated as\cite{Mat,Mat2,Mat3}
\begin{align}
\kappa_{x^{\prime}y^{\prime}}=\frac{k_B^2T}{(2\pi)^2\hbar}\sum_n\int_{\rm BZ}c_2(\rho_n)B_{k_x^{\prime}k_y^{\prime}}^n{\rm d}k_x^{\prime}{\rm d}k_y^{\prime},
\end{align}
with the sum running over the three bands and the integral is over the first Brillouin zone. $\rho_n=1/(\exp(E_n/k_BT)-1)$ is the Bose distribution with $E_n$ as the $n$th eigenvalue. $c_2$ is given by
\begin{align}
c_2(\rho_n)=(1+\rho_n)(\rm{ln}\frac{1+\rho_n}{\rho_n})^2-(\rm{ln}\rho_n)^2-2Li_2(-\rho_n),
\end{align}
where $\rm{Li}_2(x)$ is the polylogarithm function of order $2$.

As shown in Fig. \ref{Fig4}(a) and \ref{Fig4}(b), the thermal Hall conductivity are always negative and positive for the phases $(-1,0,1)$ and $(1,0,-1)$ respectively, with the sign changes happen at the phase boundaries, whose conductivity are always zero. The sign structure of the thermal conductivity stems from the sign structure of the Berry curvatures of the magnon bands. As for the phase $(-1,0,1)$, the negative and positive Berry curvatures of the first and third bands contribute to the Chern numbers -1 and 1 respectively. However, in the integral of the thermal conductivity, there is a weight $c_2(\rho_n)\ge0$ which decreases monotonically with energy\cite{Cao}, thus the negative contribution from the first band is greater than the positive contribution from the third band, which leads to the negative conductivity. For the phase $(1,0,-1)$, the situation is reversed and thus its conductivity is positive. We have checked that, the contributions from the second band are nearly zero for both phases. From the physical angle, the lowest magnon band will be thermally occupied maximally at low temperatures, and then dominate the thermal conductivity.

We note that, the phase $(-1,0,1)$ with negative thermal conductivity resides around $\psi=0.6\pi$ and $\psi=1.6\pi$ in the phase diagram, where the Kitaev interaction is relatively large with $\left|K\right|\approx0.95$. Moreover, the width of the phase increases with the strength of DMI. These phenomena will be helpful to distinguish the relative strength of the isotropic Heisenberg interaction, the Kitaev interaction and the DMI.

\section{Summary}\label{sectoin5}
We have studied the topological magnon excitations and related thermal Hall conductivity in a kagome magnet exhibiting Heisenberg, Kitaev and DM interactions exposed to a magnetic field. We consider a strong enough filed to bring the system into fully polarized paramagnetic phase. We find that the magnon bands carry nontrivial Chern numbers in the full region of the phase diagram. Furthermore, there are phase transitions relate two topological phases with opposite Chern numbers, which lead to the sign changes of the thermal Hall conductivity. In the phase with negative thermal conductivity, the Kitaev interaction is relatively large and the width of the phase increases with the strength of DMI. Since the effects of magnon-magnon interactions are suppressed by the exchange scale divided by the applied field strength, the paramagnetic phase here is suitable for the exploration of related physics. Thus we believe that the study here will contribute to the understanding of related compounds.

\section*{Acknowledgements}
We thank Changle Liu for helpful discussions. This work was supported by the National Natural
Science Foundation of China (Grant NO. 12104407) and the Natural Science Foundation of Zhejiang Province (Grant NO. LQ20A040004).

\section*{Appendix: magnon Hamiltonian matrix}
We denote the directions of spins ${\bf S}_{i=1,2,3}$ by their polar angles $\theta_{1,2,3}$ and azimuthal angles $\phi_{1,2,3}$ in the global frame. The HP transformation for a spin in its local frame reads
\begin{align}
S_x^0=&\frac{\sqrt{2S}}{2}(a+a^\dagger),\\
S_y^0=&\frac{\sqrt{2S}}{2{\rm i}}(a-a^\dagger),\\
S_z^0=&S-a^\dagger a,
\end{align}
where $a^\dagger$ and $a$ are the magnon creation and annihilation operators respectively, which obey the boson commutation rules.
Then by multiplying a rotation matrix we get the HP transformation for spin ${\bf S}_i$ in the global frame
\begin{align}
\left(
\begin{array}{c}
S_{ix}\\[8pt]
S_{iy}\\[8pt]
S_{iz}
\end{array}
\right)=\left(
\begin{array}{ccc}
\cos\theta_i\cos\phi_i & -\sin\phi_i & \sin\theta_i\cos\phi_i\\[8pt]
\cos\theta_i\sin\phi_i & \cos\phi_i & \sin\theta_i\sin\phi_i\\[8pt]
-\sin\theta_i & 0 & \cos\theta_i
\end{array}
\right)
\left(
\begin{array}{c}
S_{ix}^0\\[8pt]
S_{iy}^0\\[8pt]
S_{iz}^0
\end{array}
\right).
\end{align}
Substituting $S_{ix,y,z}$ into the Hamiltonian (\ref{model}) and then do the Fourier transformation, we get the quadratic Hamiltonian in momentum space
\begin{align}
H=\frac{1}{2}\sum_{\bm{k}^\prime}\Psi_{\bm{k}^\prime}^\dagger h(\bm{k}^\prime)\Psi_{\bm{k}^\prime},
\end{align}
where $\Psi_{\bm{k}^\prime}^\dagger=(a_{1\bm{k}^\prime}^\dagger, a_{2\bm{k}^\prime}^\dagger, a_{3\bm{k}^\prime}^\dagger, a_{1-\bm{k}^\prime}, a_{2-\bm{k}^\prime}, a_{3-\bm{k}^\prime})$. The magnon Hamiltonian matrix is
\begin{align}
h(\bm{k}^\prime)=\left(
\begin{array}{cc}
A_{\bm{k}^\prime}&B_{\bm{k}^\prime}^\dagger\\[4pt]
B_{\bm{k}^\prime}&A_{\bm{k}^\prime}^*
\end{array}
\right)S,
\end{align}
with $A_{\bm{k}^\prime}$ and $B_{\bm{k}^\prime}$ $3\times3$ matrices. Their elements are as follows
\begin{align}
A_{11}=&-2J(c_z+e_z)-2K(c_8+e_7)\notag\\
&-2\frac{D}{\sqrt{3}}(c_D+e_D)+\frac{h}{\sqrt{3}}h_1,\\
A_{22}=&-2J(c_z+d_z)-2K(c_8+d_6)\notag\\
&-2\frac{D}{\sqrt{3}}(c_D+d_D)+\frac{h}{\sqrt{3}}h_2,\\
A_{33}=&-2J(d_z+e_z)-2K(d_6+e_7)\notag\\
&-2\frac{D}{\sqrt{3}}(d_D+e_D)+\frac{h}{\sqrt{3}}h_3,
\end{align}

\begin{align}
A_{12}=&[J(c_x+c_y-{\rm i}c_{xy}+{\rm i}c_{yx})+Kc_3+\frac{D}{\sqrt{3}}c_B^*]\cos \bm{k}^\prime\cdot\bm{\delta}_1,\\
A_{21}=&A_{12}^*,\\
A_{13}=&[J(e_x+e_y+{\rm i}e_{xy}-{\rm i}e_{yx})\notag\\
&+K(e_2+e_5+{\rm i}e_{10}-{\rm i}e_{12})+\frac{D}{\sqrt{3}}e_B]\cos \bm{k}^\prime\cdot\bm{\delta}_3,\\
A_{31}=&A_{13}^*,\\
A_{23}=&[J(d_x+d_y-{\rm i}d_{xy}+{\rm i}d_{yx})\notag\\
&+K(d_1+d_4-{\rm i}d_9+{\rm i}d_{11})+\frac{D}{\sqrt{3}}d_B^*]\cos \bm{k}^\prime\cdot\bm{\delta}_2,\\
A_{32}=&A_{23}^*,
\end{align}

\begin{align}
B_{12}=&B_{21}=[J(c_x-c_y-{\rm i}c_{xy}-{\rm i}c_{yx})\notag\\
&+Kc_3+\frac{D}{\sqrt{3}}c_A]\cos \bm{k}^\prime\cdot\bm{\delta}_1,\\
B_{13}=&B_{31}=[J(e_x-e_y-{\rm i}e_{xy}-{\rm i}e_{yx})\notag\\
&+K(e_2-e_5-{\rm i}e_{10}-{\rm i}e_{12})+\frac{D}{\sqrt{3}}e_A]\cos \bm{k}^\prime\cdot\bm{\delta}_3,\\
B_{23}=&B_{32}=[J(d_x-d_y-{\rm i}d_{xy}-{\rm i}d_{yx})\notag\\
&+K(d_1-d_4-{\rm i}d_9-{\rm i}d_{11})+\frac{D}{\sqrt{3}}d_A]\cos \bm{k}^\prime\cdot\bm{\delta}_2,\\
B_{11}=&B_{22}=B_{33}=0,
\end{align}
where
\begin{align}
c_1=&\cos\theta_1\cos\phi_1\cos\theta_2\cos\phi_2,\\
c_2=&\cos\theta_1\sin\phi_1\cos\theta_2\sin\phi_2,\\
c_3=&\sin\theta_1\sin\theta_2,\\
c_4=&\sin\phi_1\sin\phi_2,\\
c_5=&\cos\phi_1\cos\phi_2,\\
c_6=&\sin\theta_1\cos\phi_1\sin\theta_2\cos\phi_2,\\
c_7=&\sin\theta_1\sin\phi_1\sin\theta_2\sin\phi_2,\\
c_8=&\cos\theta_1\cos\theta_2,\\
c_9=&-\cos\theta_1\cos\phi_1\sin\phi_2,\\
c_{10}=&\cos\theta_1\sin\phi_1\cos\phi_2,\\
c_{11}=&-\sin\phi_1\cos\theta_2\cos\phi_2,\\
c_{12}=&\cos\phi_1\cos\theta_2\sin\phi_2,
\end{align}

\begin{align}
c_x=&c_1+c_2+c_3,\\
c_y=&c_4+c_5,\\
c_z=&c_6+c_7+c_8,\\
c_{xy}=&c_9+c_{10},\\
c_{yx}=&c_{11}+c_{12},
\end{align}

\begin{align}
c_{13}=&\sin\theta_1\cos\theta_2\sin\phi_2,\\
c_{14}=&-\cos\theta_1\sin\phi_1\sin\theta_2,\\
c_{15}=&\sin\theta_1\sin\phi_1\cos\theta_2,\\
c_{16}=&-\cos\theta_1\sin\theta_2\sin\phi_2,\\
c_{17}=&-\cos\phi_1\sin\theta_2,\\
c_{18}=&\sin\theta_1\cos\phi_2,\\
c_{19}=&\cos\theta_1\cos\phi_1\sin\theta_2,\\
c_{20}=&-\sin\theta_1\cos\theta_2\cos\phi_2,\\
c_{21}=&\cos\theta_1\sin\theta_2\cos\phi_2,\\
c_{22}=&-\sin\theta_1\cos\phi_1\cos\theta_2,\\
c_{23}=&\sin\theta_1\sin\phi_2,
\end{align}

\begin{align}
c_{24}=&-\sin\phi_1\sin\theta_2,\\
c_{25}=&\cos\theta_1\cos\phi_1\cos\theta_2\sin\phi_2,\\
c_{26}=&-\cos\theta_1\sin\phi_1\cos\theta_2\cos\phi_2,\\
c_{27}=&\cos\phi_1\sin\phi_2,\\
c_{28}=&-\sin\phi_1\cos\phi_2,\\
c_{29}=&\sin\theta_1\cos\phi_1\sin\theta_2\sin\phi_2,\\
c_{30}=&-\sin\theta_1\sin\phi_1\sin\theta_2\cos\phi_2,\\
c_{31}=&\cos\theta_1\cos\phi_1\cos\phi_2,\\
c_{32}=&\cos\theta_1\sin\phi_1\sin\phi_2,\\
c_{33}=&-\sin\phi_1\cos\theta_2\sin\phi_2,\\
c_{34}=&-\cos\phi_1\cos\theta_2\cos\phi_2,
\end{align}

\begin{align}
c_{xxx}=&c_{13}+c_{14},\\
c_{xzz}=&c_{15}+c_{16},\\
c_{yxx}=&c_{19}+c_{20},\\
c_{yzz}=&c_{21}+c_{22},\\
c_{zxx}=&c_{25}+c_{26},\\
c_{zyy}=&c_{27}+c_{28},\\
c_{zzz}=&c_{29}+c_{30},\\
c_{zxy}=&c_{31}+c_{32},\\
c_{zyx}=&c_{33}+c_{34},
\end{align}

\begin{align}
c_A=&c_{xxx}+c_{yxx}+c_{zxx}-c_{zyy}\notag\\
&-{\rm i}(c_{17}+c_{24}+c_{zyx}+c_{18}+c_{23}+c_{zxy}),\\
c_B=&c_{xxx}+c_{yxx}+c_{zxx}+c_{zyy}\notag\\
&-{\rm i}(c_{17}+c_{24}+c_{zyx}-c_{18}-c_{23}-c_{zxy}),\\
c_D=&c_{xzz}+c_{yzz}+c_{zzz},
\end{align}

and change the corresponding subscripts in $\theta_{1,2}$ and $\phi_{1,2}$ to $\theta_{2,3}$ and $\phi_{2,3}$, we get the corresponding expressions for $d_i$ with
\begin{align}
i=&1-34,x,y,z,xy,yx,xxx,xzz,yxx,yzz,\notag\\
&zxx,zyy,zzz,zxy,zyx,A,B,D.
\end{align}
Similarly, change $\theta_{1,2}$ and $\phi_{1,2}$ to $\theta_{3,1}$ and $\phi_{3,1}$, we get the corresponding expressions for $e_i$. And we have
\begin{align}
h_1=\sin\theta_1\cos\phi_1+\sin\theta_1\sin\phi_1+\cos\theta_1,\\
h_2=\sin\theta_2\cos\phi_2+\sin\theta_2\sin\phi_2+\cos\theta_2,\\
h_3=\sin\theta_3\cos\phi_3+\sin\theta_3\sin\phi_3+\cos\theta_3,
\end{align}
Note that for the polarized phase here, we have $\theta_1=\theta_2=\theta_3=\arctan\sqrt{2}$ and $\phi_1=\phi_2=\phi_3=\pi/4$. The vectors $\bm{\delta}_{1,2,3}$ are the NN vectors of the kagome lattice with $\bm{\delta}_1=(1/2,0)$, $\bm{\delta}_2=(-1,\sqrt{3})/4$, $\bm{\delta}_3=(-1,-\sqrt{3})/4$, which are defined in the new 2D frame $x^\prime y^\prime$. Note that to get the eigenvalues and eigenvectors of bosonic quadratic Hamiltonian, we need to diagonalize the matrix $I_-h(\bm{k}^\prime)$ instead of $h(\bm{k}^\prime)$, where
\begin{align}
I_-=\left(
\begin{array}{cc}
I&0\\[4pt]
0&-I
\end{array}
\right),
\end{align}
with $I$ the $3\times3$ identity matrix .

\bibliography{Reference}
%\begin{thebibliography}{99}
%\bibitem{Kane1}  M. Z. Hasan and C. L. Kane, Rev. Mod. Phys. \textbf{82}, 3045 (2010).
%\bibitem{SCZhang}  X.-L. Qi and S.-C. Zhang, Rev. Mod. Phys. \textbf{83}, 1057 (2011).
%\bibitem{Kane2} F. Zhang, C. L. Kane, and E. J. Mele, Phys. Rev. Lett. \textbf{110}, 046404 (2013).
%
%\end{thebibliography}

\end{document}